\documentclass[preprint,12pt,3p,times]{elsarticle}
\usepackage{setspace}
\doublespace

\usepackage{amssymb}
\usepackage{graphicx}
\graphicspath{{figures/}}
\usepackage{amsmath}
\usepackage{subfigure}
\usepackage{geometry}
\usepackage{subfig}
\usepackage{caption}
\usepackage{multirow}
\usepackage{booktabs}
\usepackage{tabularx}
\usepackage{bm}
\usepackage{color}

\journal{}

\begin{document}

\begin{frontmatter}



\title{Simulation of magnetic hyperthermia cancer treatment near a blood vessel}


\author[label1]{Qian Jiang}
\author[label1,label2]{Feng Ren}
\author[label1]{Chenglei Wang}
\author[label1]{Zhaokun Wang}
\author[label3]{Gholamreza Kefayati}
\author[label4]{Sasa Kenjeres}
\author[label5]{Kambiz Vafai}
\author[label1]{Yang Liu}
\author[label1]{Hui Tang\corref{cor1}}
\cortext[cor1]{Email: h.tang@polyu.edu.hk}

\address[label1]{Department of Mechanical Engineering, The Hong Kong Polytechnic University, Hong Kong, China}
\address[label2]{School of Marine Science and Technology, Northwestern Polytechnical University, Xi’an, Shaanxi 710072, China}
\address[label3]{School of Engineering, University of Tasmania, Hobart 7001, Tasmania, Australia}
\address[label4]{Transport Phenomena Section, Department of Chemical Engineering, Faculty of Applied Sciences and J. M. Burgers Center for Fluid Mechanics, Delft University of Technology, Van der Maasweg 9, Delft 2629 HZ, The Netherlands}
\address[label5]{Mechanical Engineering Department, University of California, Riverside, California 92521, USA}

\begin{abstract}
In this study, we conduct a study on magnetic hyperthermia treatment when a vessel is located near the tumor. The holistic framework is established to solve the process of tumor treatment. The interstitial tissue fluid, MNP distribution, temperature profile, and nanofluids are involved in the simulation. The study evaluates the cancer treatment efficacy by cumulative-equivalent-minutes-at-43$^\circ$C (CEM43), a widely accepted thermal dose. The influence of the nearby blood vessel is investigated, and parameter studies about the distance to the tumor, the width of the blood vessel, and the vessel direction are also conducted. After that, the effects of the fluid-structure interaction of moving vessel boundaries and blood rheology are discussed. The results demonstrate the cooling effect of a nearby blood vessel, and such effect reduces with the augment of the distance between the tumor and blood vessel. The combination of downward gravity and the cool effect from the lower horizontal vessel leads to the best performance with 97.77\% ablation in the tumor and 0.87\% injury in healthy tissue at distance $d=4$ mm, but the cases of the vertical vessel are relatively poor. The vessel width and blood rheology affect the treatment by velocity gradient near the vessel wall. Additionally, the moving boundary almost has no impact on treatment efficacy. The simulation tool has been well-validated and the outcomes can provide a useful reference to magnetic hyperthermia treatment.

\end{abstract}




\end{frontmatter}


\section{Introduction}
Magnetic hyperthermia cancer treatment emerges in recent years in which tumor tissues are locally heated to approximately above $43^{\circ}$C \cite{sharma2019nanoparticles,vilas2020magnetic,ma2019theoretical}, by injecting the magnetic nanoparticles (MNPs) into tumor tissue region and exposing them to the high frequency alternating magnetic field (AMF) to locally heat the tumor cells to the appropriate temperature to ablate them \cite{perigo2015fundamentals,jose2020magnetic}. With the potential of only heating the tumor cells to death but protecting the surrounding healthy tissue, hyperthermia can effectively avoid side effects caused by conventional treatment methods and therefore alleviate suffering. Magnetite ($\text{Fe}_3\text{O}_4$) is the most popular MNPs candidate in many studies for its favorable magnetic properties and low toxicity \cite{karponis2016arsenal, kosari2021transport, chang2018biologically}. The amount of heat from MNP is defined by Rosensweig's model, which is highly related to the strength and frequency of AMF \cite{suto2009heat,rosensweig2002heating}.

Accurately predicting the magnetic hyperthermia treatment efficacy is still a challenge \cite{raouf2020review}. There already being massive numerical efforts on it, but a holistic simulation framework involving all the main factors is much rare. Pennes's bio-beat transfer equation (PBHTE), proposed by Pennes in 1948, is the most acceptable simulation model for magnetic hyperthermia treatment, based on thermal energy balance with consideration with heat convection of blood perfusion and heat generation induced by MNPs \cite{singh2020computational, mahmoudi2018magnetic}. However, since PBHTE only simulates the heat transfer process without considering other important factors, this simulation tool needs further improvement.

Mass transfer of MNP has been combined with PBHTE in some studies \cite{dahaghin2021numerical, golneshan2011diffusion}, which has also been applied in many related studies \cite{nicholson2001diffusion,roustaei2022effect,ooi2017mass}. Soltani et al. \cite{soltani2020effects} investigated the influence of MNP transfer and concluded that diffusion of MNP decreases the maximum temperature but expends the ablation region in a solid tumor. Usually, MNP distribution can be controlled at the beginning if the injection flow rate is slow enough as mentioned in the experiment \cite{salloum2008controlling}.

Several studies also involve interstitial tissue flow in simulation, solving by Brinkman equation or Darcy equation \cite{tang2018impact, tang2020effect, erbertseder2012coupled, astefanoaei2016thermofluid, zakariapour2017numerical}. Tang et al. \cite{tang2023backflow} conducted the study by considering the factors of interstitial tissue flow, heat and mass transfer, which is a really rare case. Results indicate the significant influence of interstitial flow field on the treatment efficacy. As mentioned in \cite{pedersen2007effects}, Navier-Stokes is more suitable to simulate the motion of the interstitial flow field but barely appears in the related studies. 

The existence of a nearby blood vessel has a great influence on the distribution of heat and the degree of damage  \cite{alamiri2014fluid, gheflati2020computational}. Blood is a complex mixture of erythrocytes, leukocytes, thrombocytes, iron oxides, and other components suspended in plasma \cite{tzirtzilakis2005mathematical,wang2011lattice}, which makes the shear rate of blood fluid not linearly related to the viscosity. Blood rheology can be regarded as a shear-rate-dependent non-Newtonian fluid \cite{cho1991effects}. In addition, induced by blood pulsation and deformation, the vessel wall is regarded as a moving boundary, with both longitudinal movement and diameter change \cite{cinthio2006longitudinal}.

Tumor damage is a collaboration result of temperature and time \cite{dewhirst2003basic}. CEM43 is the equilibrium accumulated exposure time at $43^\circ C$, which can deal with the non-uniform spatially and transient temporally situation during magnetic hyperthermia treatment \cite{spirou2018magnetic,kandala2018temperature, singh2020computational, sapareto1984thermal}.

Therefore, we build a holistic simulation framework for the simulation of magnetic hyperthermia treatment, where we consider the factors of interstitial flow, heat and mass transfer process, as well as important external forces, by using Navier-Stokes equation on interstitial flow field and blood flow field, using variant PBHTE on temperature field, and using concentration equation on MNP mass field. With this framework, we first investigate the influence of nearby blood vessel, and then conduct a series of parameteric studies, including the influence of gravity, distance to the tumor, the width, and the direction. Finally, the influence of moving vessel walls and blood rheology are also involved. Results of this study will provide meaningful suggestions on magnetic hyperthermia treatment.

\section{Methodology}
\subsection{Physical model}
\begin{figure}
    \centering
    \includegraphics[width=0.6\textwidth]{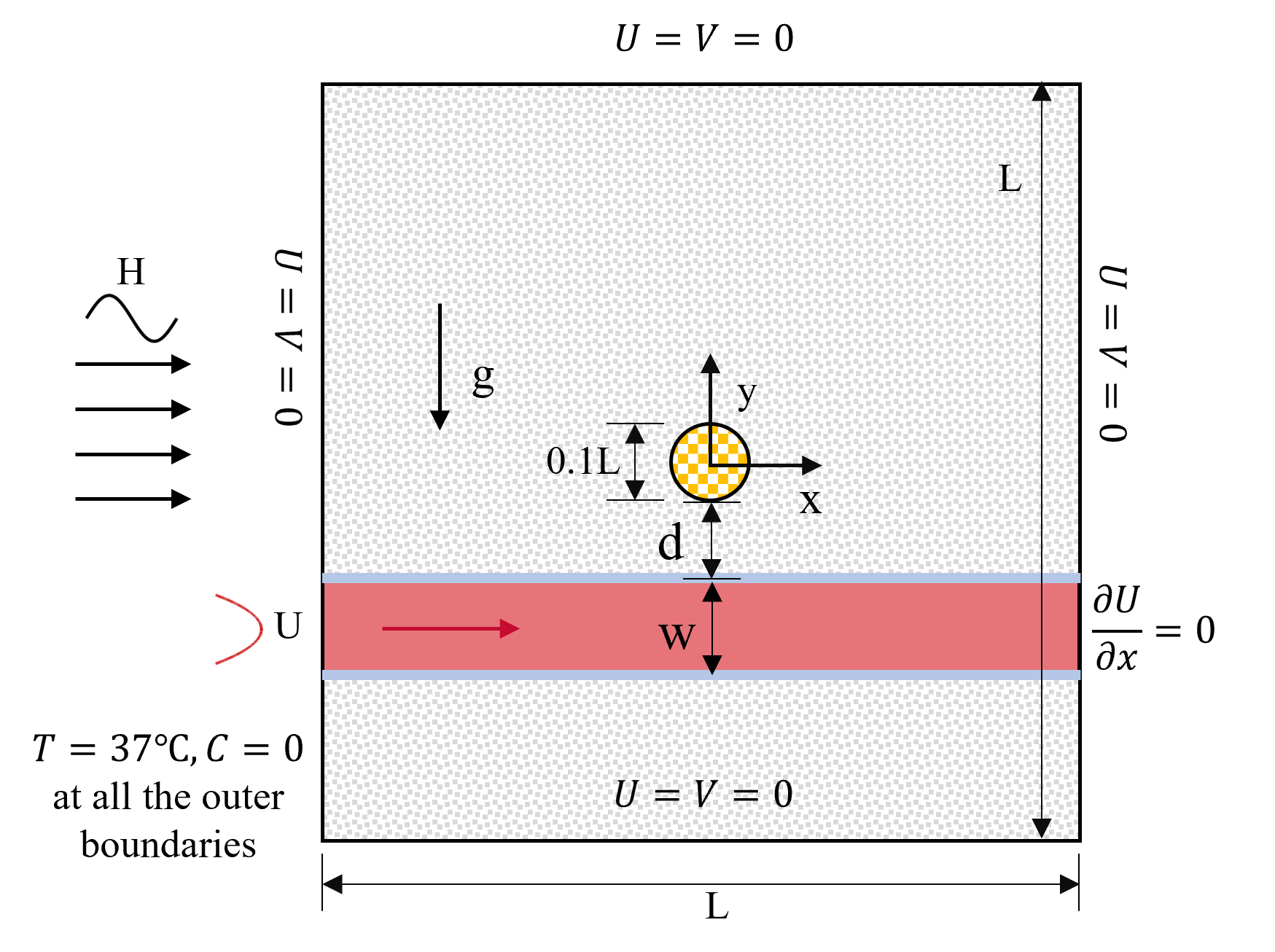}
    \caption{Schematic of magnetic hyperthermia in this study. The orange region denotes the tumor, while the gray region represents the healthy tissue. Pink channel denotes the blood vessel, and the blue lines are the vessel walls.}
    \label{FigSchematicVessel}
\end{figure}

The schematic of this problem is presented as Figure \ref{FigSchematicVessel}, where a simplified circular tumor is located at the center of the square block. The diameter of the tumor is assumed 10 mm, which is one-tenth of the length of the whole block. Both tumor and healthy tissue are treated as porous media but with different permeability. A straight blood vessel traverses the healthy tissue at the place near the tumor with the parabolic velocity profile at the inlet and Neumann boundary at the outlet. Apart from them, other outer boundaries are stationary. The vessel walls are considered moving when involving the fluid-structure interaction effect, otherwise, they are fixed. Constant zero volume fraction and core temperature ($37^\circ$C) are namely adopted for the MNP and temperature boundaries. At the beginning of treatment, MNPs are uniformly distributed with 0.01 volume fraction in the tumor region, while the velocity of fluid is stationary and temperature is $37^\circ$C.

\subsection{Governing equation}
The governing equations of this multi-physics problem is given by

\begin{subequations}
    \begin{equation}
        \nabla\cdot\mathbf{u}=0
        \label{EqnContinuity}
    \end{equation}
    \begin{equation}
        \frac{\partial\mathbf{u}}{\partial t}+(\mathbf{u}\cdot\nabla)\Big(\frac{\mathbf{u}}{\phi}\Big)=-\frac{1}{\rho_{nf}}\nabla(\phi p)+\upsilon_{nf}\nabla^2\mathbf{u}+\frac{\mathbf{F}}{\rho_{nf}}
        \label{EqnMomentum}
    \end{equation}
    \begin{equation}
        \sigma\frac{\partial T}{\partial t}+\mathbf{u}\cdot\nabla T=\alpha_e\nabla^2T+\frac{1}{(\rho c_p)_{nf}}\dot{m}_bc_{pb}(T_b-T)+C\frac{Q}{(\rho c_p)_{nf}}
        \label{EqnEnergy}
    \end{equation}
    \begin{equation}
        \phi\frac{\partial C}{\partial t}+\mathbf{u}\cdot\nabla C=D_e\nabla^2C
        \label{EqnConcentration}
    \end{equation}
    \label{EqnGoverning}
\end{subequations}
where $\nabla\equiv\frac{\partial}{\partial x}\bm{i}+\frac{\partial}{\partial y}\bm{j}$. 

Equation \ref{EqnGoverning} namely lists continuity equation, momentum equation, energy equation, and concentration equation. In this holistic simulation framework, they are solved by multiple-relaxation-time Lattice Boltzmann Method (MRT-LBM), with the D2Q9 scheme on fluid field while D2Q5 scheme on temperature and concentration fields \cite{liu2014multiple}. $\mathbf{u}$, $T$ and $C$ denote the fluid velocity vector (u,v) along x and y directions, temperature and MNPs volume fraction, respectively. This set of equations are adaptable in tumor tissue, healthy tissue and the blood vessel region. $\phi$ means the porosity of tissue, and $p$ means pressure. $\rho$, $c_p$, $\upsilon$ are fluid density, specific heat and kinetic viscosity. Since blood is a complex mixture, blood rheology is considered in part of the cases for influence study. The Carreua-Yasuda Model is applied, and it is detailed illustrated in the following section. Coefficient $\sigma=[(\rho c_p)_{nf}\phi+(\rho c_p)_s(1-\phi)]/(\rho c_p)_{nf}$. $k_e=k_{nf}\phi+k_s(1-\phi)$ \cite{mehmood2017numerical, hussain2018numerical} is the effective thermal conductivity of porous media, and effective thermal diffusivity is defined as $\alpha_e=k_e/(\rho c_p)_{nf}$. $D_e=\phi D$ \cite{hussain2018numerical} is the effective concentration diffusivity of MNPs in porous media, and $D$ is the concentration diffusivity in pure fluid. Specifically, subscript “$nf$” denotes properties of nanofluid or modified tissue that is a mixture combining the tissue and the injected MNPs, while “$b$” denotes properties of blood, and “$s$” denotes property of tissue structure. 

$\mathbf{F}$ in the last term of momentum equation (see Equation \ref{EqnMomentum}) is the total external body force, including the resistance force from porous media, gravity force caused by temperature and concentration gradient, Lorentz force induced by AMF, and the body force of the immersed boundary, as \cite{liu2014multiple}

\begin{equation}
        \mathbf{F}=\frac{1}{\rho_{nf}}(-\frac{\varphi\mu_{nf}}{K}\mathbf{u}-\frac{\varphi\rho_{nf} F_\varphi}{\sqrt{K}}|\mathbf{u}|\mathbf{u}+\varphi\mathbf{G}+\varphi\mathbf{F_M}+\varphi\mathbf{F_b})
    \label{EqnExterF}
\end{equation}

where the resistance force in porous media comes from the representative elementary volume (REV) scale method \cite{guo2002lattice,liu2014multiple}. $F_\phi$ in Equation \ref{EqnGoverning} denotes the Forchheimer coefficient of porous media while $K$ denotes the permeability. $F_\phi$ is only determined by the porosity $\phi$ as $F_\phi=1.75/\sqrt{150\phi^3}$, but $K$ is determined by the combination of $\phi$ and mean pore diameter $d_p$ as $K=(\phi^3d_p^2)/[150(1-\phi)^2]$. Due to the physical difference between tumor and healthy tissue, the values of permeability are not the same in their regions, and they can be separated by the subscription as $K_{tum}$ and $K_{tis}$. $|\mathbf{u}|=\sqrt{u^2+v^2}$ is the amplitude of velocity. 

With the assumption of Boussinesq approximation, buoyancy force $\mathbf{G}$ is given by \cite{liu2018multiple}

\begin{equation}
    \mathbf{G}=g[(\rho\beta_T )_{nf}(T-T_c)+(\rho\beta_C )_{nf}(C-C_c )]\mathbf{j}
    \label{EqnExterG}
\end{equation}

where $g$ is the acceleration of gravity. $(\rho\beta_T )_{nf}$ and $(\rho\beta_C )_{nf}$ are the thermal and concentration expansion of nanofluid respectively. $T_c$ and $C_c$ are namely the reference the temperature and concentration. In this study, $T_c$ equals to the core temperature of human body $37^\circ$C, and $C_c$ is zero volume fraction. $\mathbf{j}$ is the unit vector on y direction. 

$\mathbf{F_M}$ is the Lorentz force that induced by a horizontal high frequency alternating magnetic field, which is converted into a steady model as in our previous work.
\begin{equation}
    \mathbf{F_M}=-\frac{1}{2}\sigma_{nf}B_0^2v\mathbf{j}
\end{equation}
where $\sigma_{nf}$ is the electrical conductivity of nanofluid, and $B_0$ is magnetic induction amplitude, which is proportional to the magnetic field intensity amplitude $H_0$ as $B_0=\mu_0H_0$, and $\mu_0$ is magnetic permeability of vacuum. 

$\mathbf{F}_b$ is the boundary force from the immersed boundary method, which is illustrated in detail in \ref{AppendixImmersedBoundary}.

In energy equation (see Equation \ref{EqnEnergy}), the heat sink caused by blood perfusion $\dot m_b c_{pb}(T_b-T)$ and the heat source induced by the MNP exposed in alternating magnetic field $CQ$ are considered. $T_b=37^\circ$C is the temperature of the blood and $\dot m_b$ is the density flow rate of temperature-dependent perfusing blood. According to Lang \cite{lang1999impact}, $\dot m_b$ depends on local temperature $T$ as 

In healthy tissue:

\begin{equation}
    \dot m_b=
    \begin{cases}
            0.45+3.55 \exp [-{(T-45.0)^2}/{12.0}] & {T \le 45.0^{\circ}C} \\
            4.0  & {T > 45.0^{\circ}C} \\
    \end{cases}
\end{equation}

In tumor:

\begin{equation}
    \dot m_b=
    \begin{cases}
        0.833  & {T < 37.0^{\circ}C}\\
        0.8333- (T-37.0)^{4.8}/{5438.0} & {37.0 < T \le 42.0^{\circ}C}\\
        0.416 & {T > 42.0^{\circ}C}
    \end{cases}
\end{equation}

$Q$ is defined by the Rosensweig’s model \cite{rosensweig2002heating} as

\begin{equation}
    Q=\pi\mu_{0}\chi_{0}H^2_0f\frac{2\pi f\tau_{R}}{1+(2\pi f\tau_{R})^2}.
    \label{EqnQ}
\end{equation}

$H_0$ and $f$ are amplitude and frequency for external alternating magnetic field. $\chi_{0}$ denotes equilibrium susceptibility and $\tau_{R}$ denotes the effective relaxation time, which is determined by both Neel and Brownian relaxation time \cite{chang2018biologically}.

The following parameters are used to nondimensionalize the governing equation,
\begin{equation}
    \begin{split}
    X=\frac{x}{L},\;Y=\frac{y}{L},\; U=\frac{u}{U_0},\;V=\frac{v}{U_0},\;\tau=\frac{tU_0}{L},\\
    P=\frac{p}{\rho_fU_0^2},\;\theta=\frac{T-T_c}{T_h-T_c},\; \varphi=\frac{C-C_c}{C_h-C_c}
    \end{split}
\end{equation}

Then the dimensionless governing equation is given by
\begin{subequations}
    \begin{equation}
        \nabla\cdot\mathbf{U}=0
    \end{equation}
    \begin{equation}
    \begin{split}
        &\frac{\partial\mathbf{U}}{\partial\tau}+(\mathbf{U}\cdot\nabla^*)\Big(\frac{\mathbf{U}}{\phi}\Big) \\
        &=-\frac{\rho_f}{\rho_{nf}}\nabla^*(\phi P)+\frac{1}{Re}\frac{\upsilon_{nf}}{\upsilon_f}\nabla^{*2}\mathbf{U}-\phi\frac{\upsilon_{nf}}{\upsilon_f}\frac{1}{Re\cdot Da}\mathbf{U}-\phi\frac{F_\phi}{\sqrt{Da}}\sqrt{|\mathbf{U}|}\mathbf{U} \\
        &+[\phi\frac{(\rho\beta)_{nf}}{(\rho\beta)_f}\frac{\rho_f}{\rho_{nf}}Ra\frac{Re^2}{Pr}(\theta+N\varphi)-\phi\frac{\sigma_{nf}}{\sigma_n}\frac{\rho_f}{\rho_{nf}}Ha^2Pr\mathbf{U}]\mathbf{j}
    \end{split}
    \end{equation}
    \begin{equation}
        \sigma\frac{\partial \theta}{\partial\tau}+\mathbf{U}\cdot\nabla^*\theta=\frac{\alpha_e}{\alpha_f}\nabla^{*2}\theta-\frac{(\rho c_p)_f}{(\rho c_p)_{nf}}\frac{Pe}{Re\cdot Pr}\theta+\varphi\frac{(\rho c_p)_f}{(\rho c_p)_{nf}}\frac{Q_0}{Re\cdot Pr}
    \end{equation}
    \begin{equation}
        \phi\frac{\partial\varphi}{\partial\tau}+\mathbf{U}\cdot\nabla^*\varphi=\frac{\phi}{Le}\nabla^{*2}\varphi
    \end{equation}
    \label{EqnDimenLessGoverningEq3Field}
\end{subequations}

From the above dimensionless governing equations, this problem is characterized by the following dimensionless parameters:

\begin{subequations}
\begin{equation}
        Re=\frac{U_0 L}{\upsilon_f}
    \end{equation}
    \begin{equation}
        Pr=\frac{\upsilon_f}{\alpha_f}
    \end{equation}
    \begin{equation}
        Le=\frac{\alpha_f}{D}
    \end{equation}
    \begin{equation}
        Da=\frac{K}{L^2}
    \end{equation}
    \begin{equation}
        Ra=\frac{\beta_Tg(T_h-T_c)L^3}{\upsilon_f\alpha_f}
    \end{equation}
    \begin{equation}
        N=\frac{\beta_C(C_h-C_c)}{\beta_T(T_h-Tc)}
    \end{equation}
    \begin{equation}
        Ha^2=\frac{\sigma_n\mu_0^2H_0^2L^2}{\rho_f\upsilon_f}
    \end{equation}
    \begin{equation}
        Pe=\frac{\dot{m}_bc_{pb}L^2}{(\rho c_p)_f\alpha_f}
    \end{equation}
    \begin{equation}
        Q_0=\frac{QL^2(C_h-C_l)}{(\rho c_p)_f\alpha_f(T_h-T_c)}
        \label{EqnQ0}
    \end{equation}
    \label{EqnGoverningEq3FieldDimensionless}
\end{subequations}
in which, the parameters $Re$ $Pr$, $Le$, $Da$, $N$, $Ha$, $Pe$ and $Q_0$ are Reynolds number, Prandtl number, Lewis number, Darcy number, buoyancy ratio, Hartmann number, Peclet number and heat source number.

Additionally, for the reason of different permeability in tumor and tissue, the Darcy ratio is 

\begin{equation}
    R_{Da}=\frac{Da_{tum}}{Da_{tis}}=\frac{K_{tum}}{K_{tis}}
\end{equation}

The effective properties of tissue fluid should be modified by considering the influence of interspersed MNPs, and they are computed from \cite{buongiorno2006convective,gibanov2017convective}
\begin{subequations}
    \begin{equation}
        \rho_{nf}=C \rho_n+(1-C)\rho_f
    \end{equation}
    \begin{equation}
        \upsilon_{nf}=\frac{\mu_f}{\rho_{nf}(1-C)^{2.5}}
    \end{equation}
    \begin{equation}
        (\rho c_p)_{nf}=C(\rho c_p)_n+(1-C)(\rho c_p)_f
    \end{equation}
    \begin{equation}
        (\rho\beta_T)_{nf}=C(\rho\beta_T)_n+(1-C)(\rho\beta_T)_f
    \end{equation}
    \begin{equation}
        k_{nf}=k_f\frac{k_n+2k_f-2C(k_f-k_n)}{k_n+2k_f+C(k_f-k_n)}
    \end{equation}
    \begin{equation}
        \sigma_{Enf}=\sigma_{Ef}\frac{\sigma_{En}+2\sigma_{Ef}-2C(\sigma_{Ef}-\sigma_{En})}{\sigma_{En}+2\sigma_{Ef}+C(\sigma_{Ef}-\sigma_{En})}
    \end{equation}
    \label{EqnNano}
\end{subequations}
where the subscript “$f$” means pure fluid and “$n$” means nanoparticles in Equation \ref{EqnDimenLessGoverningEq3Field} and Equation \ref{EqnNano}. In this study, pure fluid represents the interstitial tissue fluid and blood flow, in which distributes the MNP $Fe_3O_4$. The kinetic and thermal proprieties of them are listed as Table \ref{TableNanofluidForPorous} \cite{zhang2008lattice,gibanov2017convective,tzirtzilakis2005mathematical}.

\begin{table*}[]\centering
    \caption{Properties of nanofluid}
    \begin{tabular}{cccc}
    \toprule
    Properties for tissue & Value & Properties for MNPs & Value  \\ \midrule
    $\rho_f (kg/m^3)$ & 1052 & $\rho_n (kg/m^3)$ & 5200  \\
    $k_f (W/mK)$ & 0.5  & $k_n (W/mK)$ & 6  \\
    $c_{pf} (J/kgK)$ & 3800 & $c_{pn} (J/kgK)$ & 670  \\
    $\beta_{Tf} (1/K)$ & $2.1\times10^{-4}$ & $\beta_{Tn} (1/K)$ & $1.3\times10^{-5}$ \\
    $\sigma_{Ef} (\Omega^{-1}\cdot m^{-1})$ & 0.7 & $\sigma_{En} (\Omega^{-1}\cdot m^{-1})$ & $2.5\times 10^4$ \\
    $\mu_f (Pa\cdot s)$ & $6.92\times 10^{-4}$ & - & -  \\ \bottomrule
    \end{tabular}
    \label{TableNanofluidForPorous}
\end{table*}

\subsection{Thermal dose}
The cumulative-equivalent-minutes-at-$43^{\circ}C$ (CEM43) model is widely accepted in thermal dose evaluating, by converting the treatment to an equivalent time on $43^{\circ}C$, as 

\begin{equation}
    CEM43=\sum_{i=1}^l{C_{EM}}^{43-T_i}\delta t
    \label{EqnCEM43}
\end{equation}
where $T_i$ is the averaged temperature in $^{\circ}C$ at the $i$th time steps, $\delta t$ represents the time interval, and $l$ denotes the total number of time steps. ${C_{EM}}$ equals to 0.5 when $T_i>43^{\circ}C$ and 0.25 otherwise \cite{sapareto1984thermal}. As CEM43 achieves 60 minutes, cells are regarded totally destroyed \cite{singh2020computational,dewhirst2003basic}. Upon this hypothesis, an ablated area ratio in tumor or surrounding healthy tissue is defined for therapeutic efficacy, i.e.
\begin{equation}
    R_{CEM43}=\frac{S_{CEM43\geq60min}}{S_{tum}}
    \label{EqnCEM43Percentage}
\end{equation}

The optimal result of $R_{CEM43}$ in tumor and healthy tissue are 1 and 0, respectively, in accordance with the expected hyperthermia treatment efficacy - totally killing the tumor cells but without destroying the healthy tissue, 

\section{Results and discussion}
\subsection{Baseline case}
The baseline case uses the properties from practical problems. Dimensionless parameters that computing from the physical case \cite{tang2020effect,zhang2008lattice,alamiri2014fluid,swartz2007interstitial,vennemann2007vivo} are listed as Table \ref{TableDimensionlessParameters}. 

\begin{table}[]\centering
    \caption{Dimensionless parameters on baseline case}
    \begin{tabular}{cc}
    \toprule
    Parameters & Value\\ \midrule
    $Re$ & $100$  \\
    $Ra_T$ & $2.05\times 10^{8}$   \\
    $Pr$ & $5.26$ \\
    $Le$ & $125.08$ \\
    $N$ & $-18.78$ \\
    $Ha$ & $4.27$3 \\
    $Q$& $572.73$ \\
    $Da_{tis}$ & $2.00\times10^{-11}$ \\
    $R_{Da}$ & $4.84$ \\
    $\phi$ & $0.26$\\ \bottomrule
    \end{tabular}
    \label{TableDimensionlessParameters}
\end{table}

\subsection{Effect of blood vessel}
\label{SecEffectVessel}
\begin{figure}
    \centering
    \includegraphics[width=\textwidth]{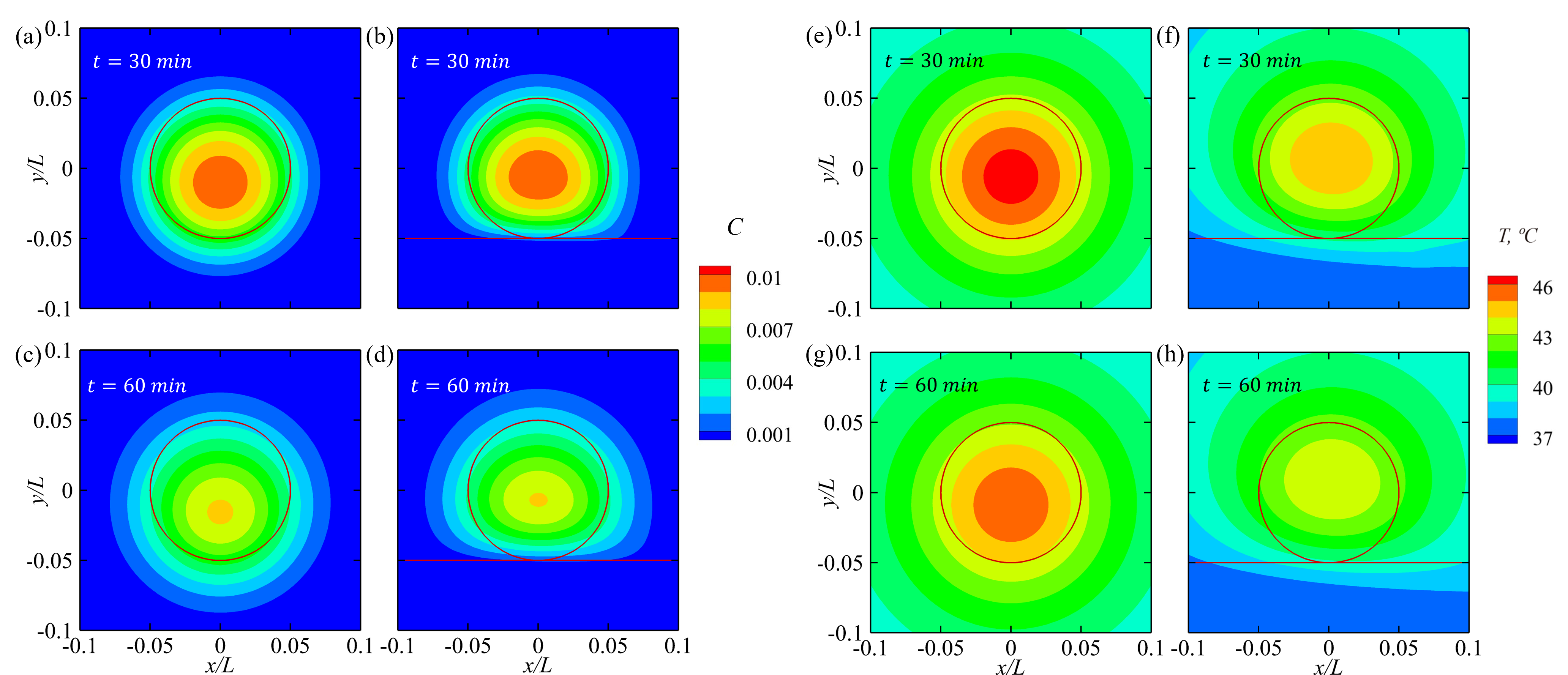}
    \caption{Effect of a nearby blood vessel on the MNP concentration and temperature at treatment time $t=30$ min and $t=60$ min. (a) $\sim$ (d) denote the MNP distribution, for no vessel case at (a) $t=30$ min and (c) $t=60$ min, and for vessel case at (b) $t=30$ min and (d) $t=60$ min. (e) $\sim$ (h) present the temperature profile for no vessel case at (a) $t=30$ min and (c) $t=60$ min, and for vessel case at (b) $t=30$ min and (d) $t=60$ min.}
    \label{FigVesselEffectTC}
\end{figure}

To explore the effect of the nearby blood vessel on the magnetic hyperthermia treatment efficacy, a vessel with $w=5mm$ and $d=0mm$ is employed for discussion. Figure \ref{FigVesselEffectTC} illustrates the effect of a nearby blood vessel on the MNP concentration and temperature. For easy comparison, two instants are taken into consideration, namely $t=30$ min and $t=60$ min. For the MNPs distribution, they diffuse obviously with time, and the flow in the blood vessel washes away the MNPs in it. Although vanishing in the vessel part, the shape of MNP distribution in healthy tissue and tumor keeps unchanged except for that near the vessel wall. For the temperature profile, the effect seems much more considerable. The maximum temperature drops with time as MNP diffusion. As Figure \ref{FigVesselEffectTC} (e), the most value of temperature exceeds $46^\circ C$, but there is only $44^\circ C \sim 45^\circ C$ in Figure \ref{FigVesselEffectTC} (f), which is a similar difference between Figure \ref{FigVesselEffectTC} (g) and (h). This means the blood vessel has a significant cooling effect on the temperature distribution of the nearby region, the came consequence as discussed in \cite{gheflati2020computational}.

\begin{figure}
    \centering
    \includegraphics[width=\textwidth]{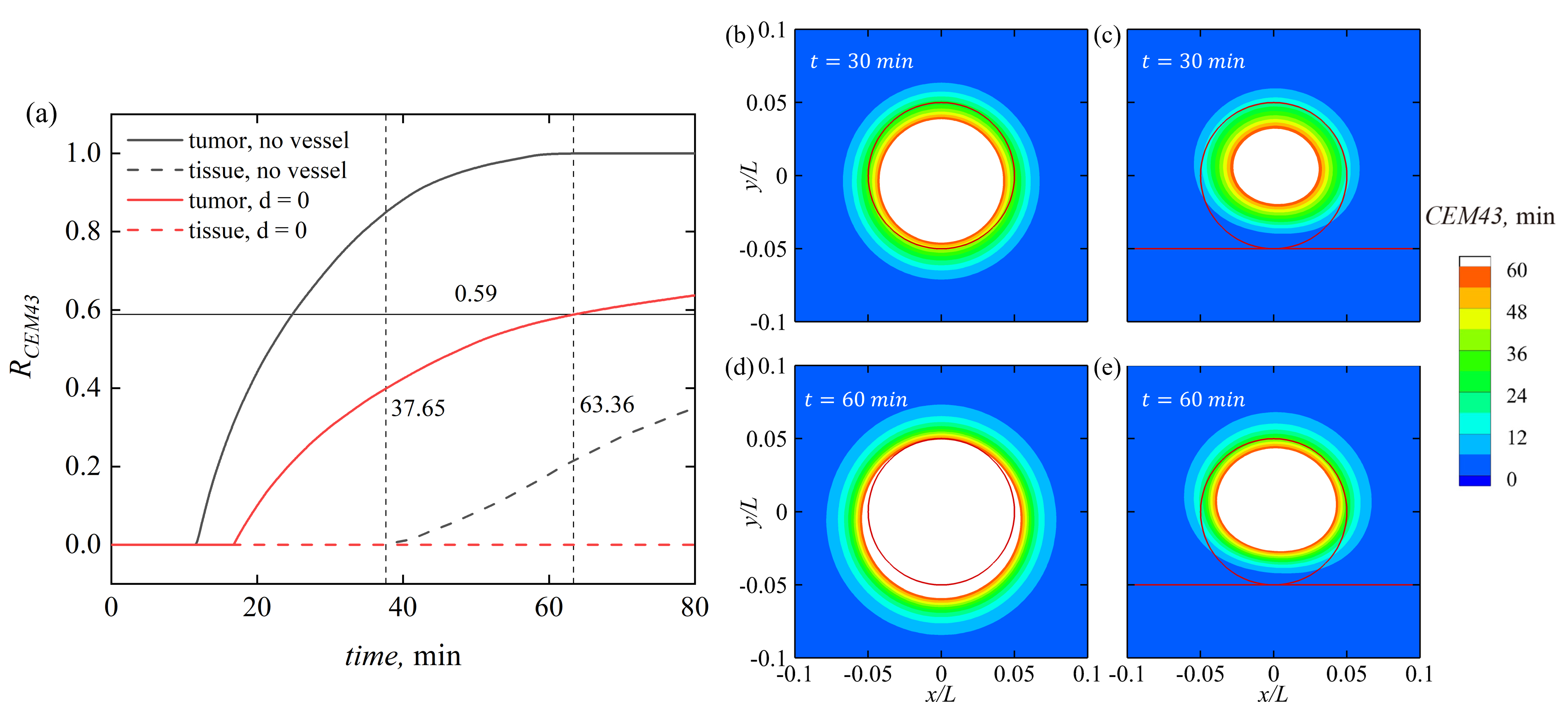}
    \caption{Effect of a nearby blood vessel on the thermal dose CEM43. (a) denotes $R_{CEM43}$ evolution in tumor and healthy tissue of cases with the nearby blood vessel or not. (b) $\sim$ (e) the CEM43 distribution, for no vessel case at (b) $t=30$ min and (d) $t=60$ min, and for vessel case at (c) $t=30$ min and (e) $t=60$ min. White regions in (b) $\sim$ (e) denote the ablation part.}
    \label{FigVesselEffectCEM43}
\end{figure}

Figure \ref{FigVesselEffectCEM43} (a) demonstrates the evolution of $R_{CEM43}$, where solid curves represent the situations of the tumor while the dashed curves are for healthy tissue. It is evident that $R_{CEM43}$ of no vessel case elevates much faster than that of vessel case. The tumor has been totally destroyed for no vessel case at $t=63.36$ min, but at the same time there are only 59\% 
tumor cells ablated for vessel case, nearly half of no vessel case. Additionally, for no vessel case, healthy tissue begins to be killed at treatment time $t=37.65$ min, but for no vessel case, still no cell is injured when the heat process lasts for 80 min. Therefore, the cooling effect of the blood vessel significantly diminishes the heating process and prolongs the treatment time. This is also reflected in the distribution of CEM43. For each instance, the ablation part is noticeably larger for no vessel case, and the whole ablation area are namely 2.36 and 2.06 times that of the vessel case. These data can be also discovered in the Table \ref{TableVesselEffectCTCEM43}. The table reveals the effect of the blood vessel on parameters in detail. As table listed, the location of $T_{max}$ marginally moves upwards and rightwards, while the $C_{max}$ moves upwards.

\begin{table}
    \centering
    \caption{Effect of a blood vessel on the parameters of magnetic hyperthermia treatment}
    \begin{tabular}{c|cc|cc}
        \bottomrule
        $t$, min & \multicolumn{2}{c|}{30} &\multicolumn{2}{c}{60} \\ \hline
        Condition & No vessel & With vessel & No vessel & With vessel   \\ 
        $T_{max}$ & 46.49 & 44.93 & 45.74 & 43.93   \\
        Location of $T_{max}$ & (0.4,0.495) & (0.5025,0.505) & (0.5,0.495) & (0.5025,0.5075)  \\
        $C_{max}\cdot10^{-3}$ & 9.64 & 9.70 & 8.17 & 8.07 \\
        Location of $C_{max}$ & (0.5,0.49) & (0.5,0.495) & (0.5,0.485) & (0.5,0.4925) \\
        $R_{CEM43}$ in tum. & 0.711 & 0.301 & 0.998 & 0.573 \\
        $R_{CEM43}$ in healthy tis. & 0 & 0 & 0.18 & 0 \\
        \toprule        
    \end{tabular}
    \label{TableVesselEffectCTCEM43}
\end{table}

\subsection{Effect of gravity}
\label{EffectGravity}
As discussed in our previous work, gravity influences treatment efficacy when a tumor is surrounded by healthy tissue. Therefore, this chapter also investigates the effect of gravity when the tumor is near a blood vessel. The cases without gravity are included for comparison with three different distances to the tumor boundary, namely $d=0$ mm, $d=2$ mm, and $d=4$ mm.

\begin{figure}
    \centering
    \includegraphics[width=\textwidth]{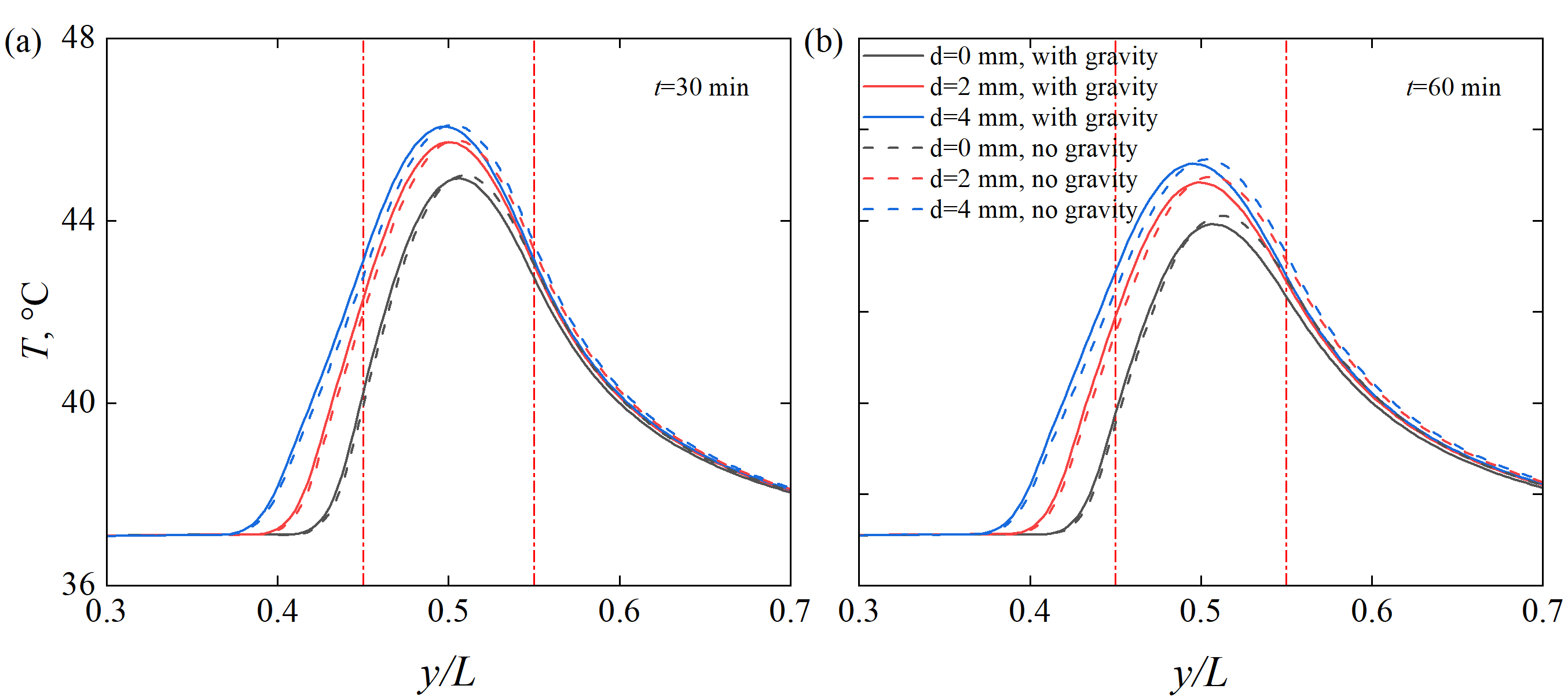}
    \caption{Effect of gravity on temperature profile along $y/L=0.5$ at treatment time (a) $t=30$ min and (b) $t=60$ min. Vertical red dash-dot lines denote the tumor boundary.}
    \label{FigGravityEffectOnT}
\end{figure}

Figure \ref{FigGravityEffectOnT} presents the temperature distribution at the vertical midline. It is not surprising to discover a downward tendency of the temperature profile and a little reduction on the maximum temperature with gravity, but this change is much weak. Although the difference enlarges with time, it is still very limited from about $0.05^\circ C$ to about $0.1^\circ C$.

\begin{table}
    \centering
    \caption{Effect of gravity on $R_{CEM43}$ at treatment time $t=60$ min}
    \begin{tabular}{cccc}
    \toprule
    $d$, mm & Parameter & No gravity & Gravity    \\ \midrule
    \multirow{2}{*}{4} & $R_{CEM43}$ in tumor, \% & 95.62 & 97.77   \\
     & $R_{CEM43}$ in healthy tissue, \% & 4.4 & 0.87   \\
    \multirow{2}{*}{2} & $R_{CEM43}$ in tumor, \% & 85.52 & 85.82   \\
     & $R_{CEM43}$ in healthy tissue, \% & 1.19 & 0.00   \\
    \multirow{2}{*}{0} & $R_{CEM43}$ in tumor, \% & 59.24 & 57.32   \\
     & $R_{CEM43}$ in healthy tissue, \% & 0.00 & 0.00   \\ 
    \bottomrule
    \end{tabular}
    \label{TableEffectGravityRCEM}
\end{table}

\begin{figure}
    \centering
    \includegraphics[width=\textwidth]{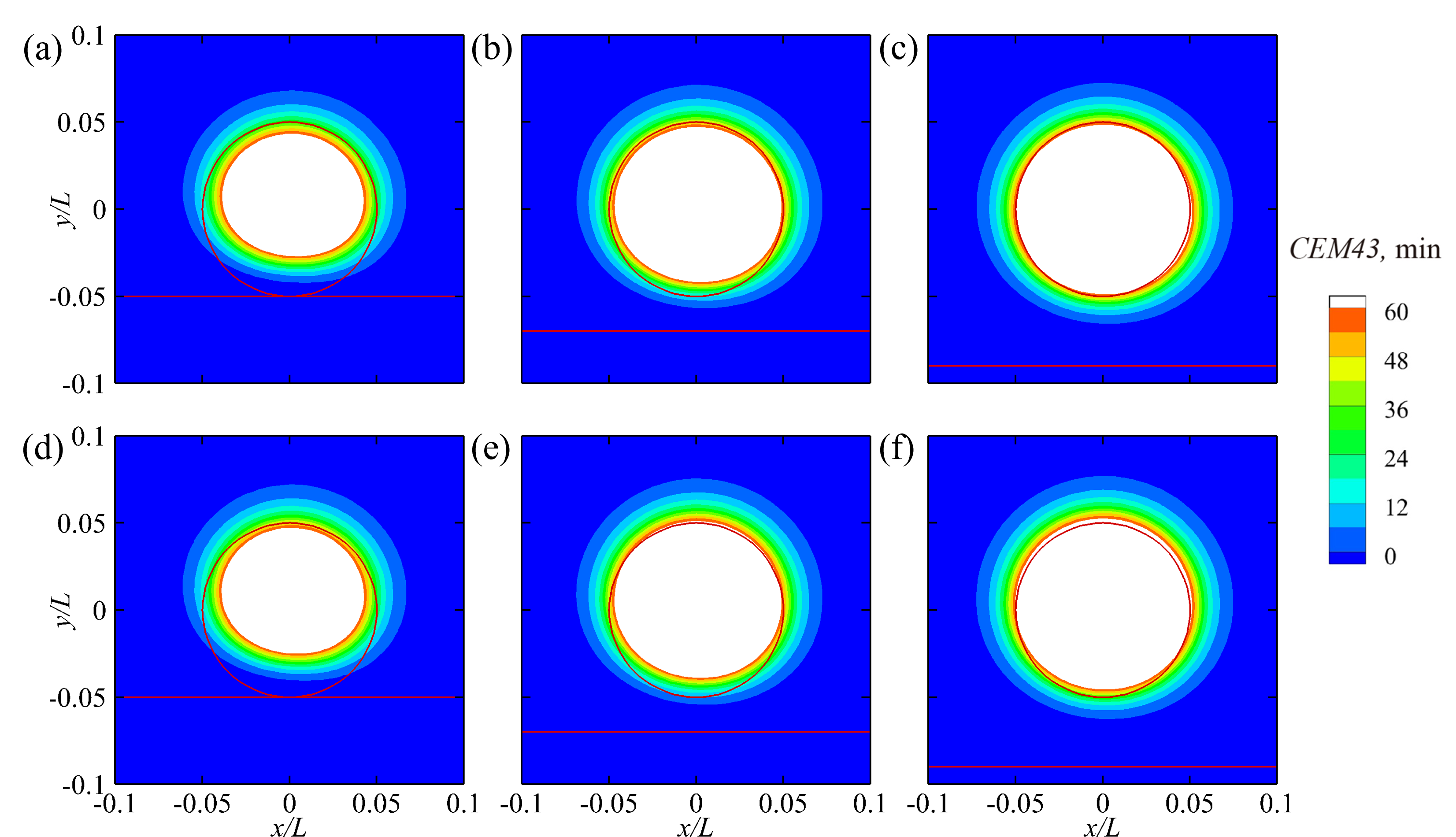}
    \caption{Effect of gravity on $CEM43$ distribution. (a) $\sim$ (c) denote the situations affected by gravity with the distance to tumor boundary (a)$d=0$ mm, (b)$d=2$ mm, and (c)$d=4$ mm. (d) $\sim$ (f) denote the situation without gravity. The red circle means the tumor boundary, and the horizontal red lines point the location of the blood vessel. White color is the ablation part.}
    \label{FigEffectGravityCEM}
\end{figure}

Table \ref{TableEffectGravityRCEM} lists the value of $R_{CEM43}$ at $t=60$ min. Generally, the existence of gravity elevates the treatment efficacy to some extent, which performs best in the case $d=4$ mm. As the results unveil, gravity leads to more ablation on the tumor but less on healthy tissue when distance equals to 4 mm, as already 97.77\% and only 0.87\%, respectively. This outcome gets pretty close to the theoretically best results, 100\% in tumor and 0\% in healthy tissue. In the case of $d=2$ mm, although $R_{CEM43}$ in tumor shows equality, there is 1.19\% injury in healthy tissue without gravity but no ablation with gravity. They are also reflected in Figure \ref{FigEffectGravityCEM}. It is not difficult to understand that gravity effect moves downwards the ablation region, which cooperates with the cooling effect of the blood vessel and performs best at the distance $d=4$ mm as Figure \ref{FigEffectGravityCEM} (c). In addition, Figure \ref{FigEffectGravityCEM} and Table \ref{TableEffectGravityRCEM} also indicates the essential influence from the distance from the upper blood vessel wall to the tumor boundary. Therefore, the investigation of distance $d$ is also discussed.

\subsection{Effect of distance}
\begin{figure}
    \centering
    \includegraphics[width=\textwidth]{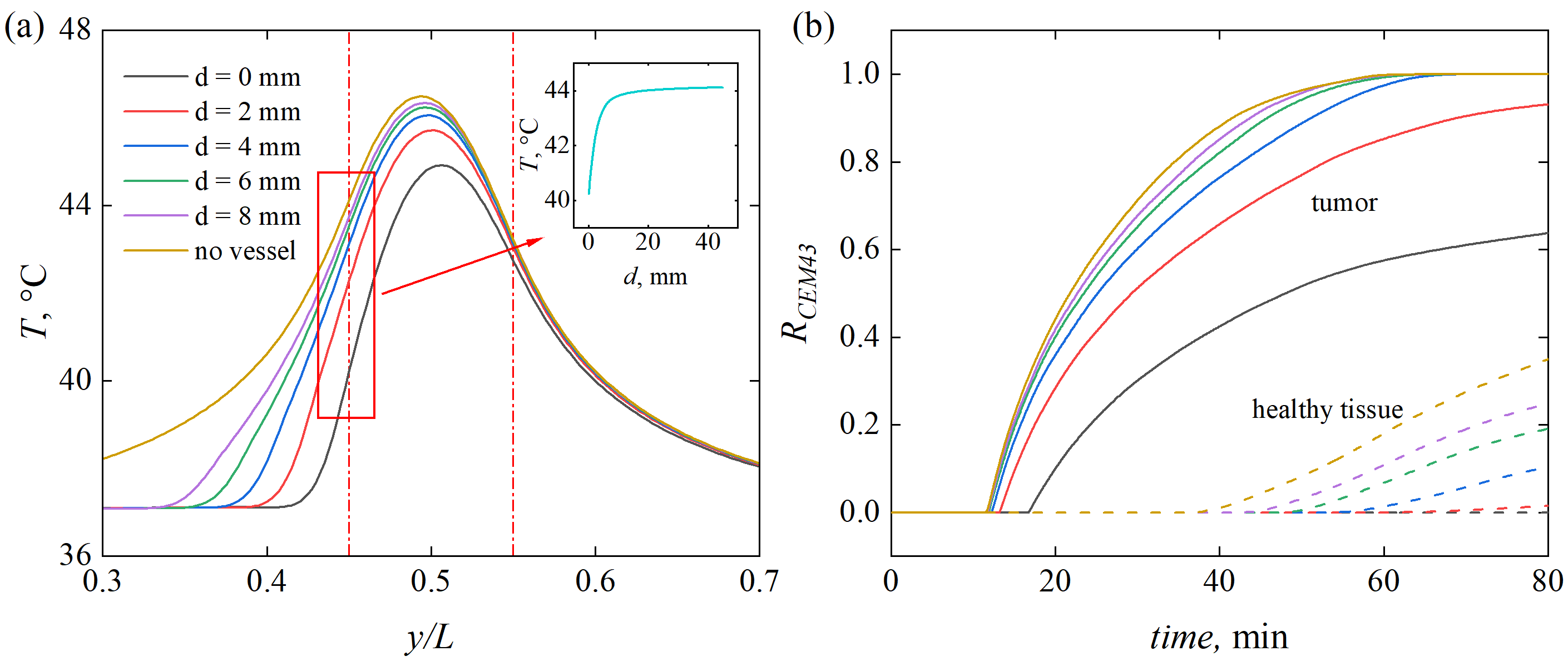}
    \caption{Effect of distance between blood vessel upper boundary to the tumor boundary $d$ at treatment time $t=30$ min. (a)The temperature profile at the midline of $x/L=0.5$, and (b) $R_{CEM43}$ evolution in tumor and healthy tissue, with the variation of distance $d$. Vertical red dash-dot lines denote the boundary of the tumor.}
    \label{FigEffectDistance}
\end{figure}

To explore the important influence of distance from the upper boundary of blood vessel to the nearest tumor boundary, six cases are employed with the variation of distance $d=$ 0 mm, 2 mm, 4 mm, 6 mm, 8 mm, and $\infty$ (no vessel case). The data reveal that when distance $d$ gradually rises from 0 mm, the temperature profile is continuously approaching to the case without the blood vessel as Figure \ref{FigEffectDistance} (a). This change is remarkable at the lower tumor boundary, where the temperature increases from $40.22^\circ C$ for $d=0$ mm to $43.75^\circ C$ for $d=8$ mm. Noticeably, the temperature at the lower tumor boundary for the no vessel case is $44.12^\circ C$, so the case with distance $d=8$ mm shows 0.8\% diffidence to that value. Not surprisingly, treatment processing slows down a lot for distance $d=0$ mm, in which only 63.77\% tumor cells are destroyed at treatment $t=80$ min. When distance $d$ augments to 2 mm, the situation gets better. There has been 93.16\% injury and 1.35\% at treatment time $t=80$ min. The cases with distance $d=$4, 6, and 8 are closer to the no vessel case, and especially, the time for totally killing the tumor cells with the distance $d=8$ mm is 65.66 min, 3.64\% longer than that of no vessel case.

\subsection{Effect of the width of blood vessel}
\begin{figure}
    \centering
    \includegraphics[width=\textwidth]{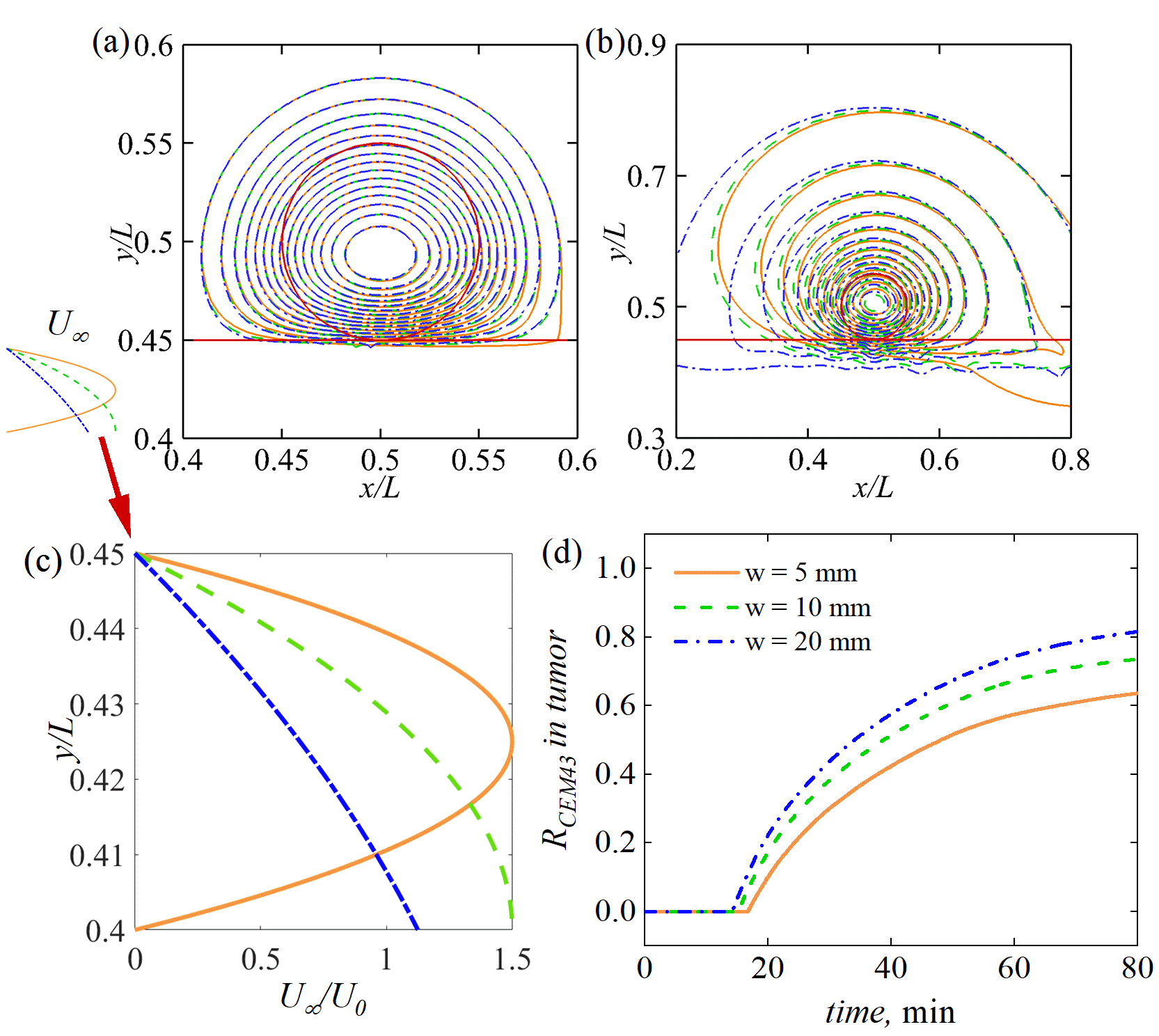}
    \caption{Effect of the width $w$ of the blood vessel on (a) MNP concentration distribution at $t=60$ min, (b) temperature distribution at $t=60$ min, and (d) the evolution of $R_{CEM43}$ in tumor. (c) presents the velocity profile at the vessel inlet. The red circle is the tumor boundary, while the horizontal red line means the upper boundary of the blood vessel. }
    \label{FigEffectWidthCEMCT}
\end{figure}
The same mean velocity at the blood vessel inlet but the different width vessel is imposed, which leads to a changeable velocity gradient near the upper boundary of the vessel. In this part, the distance from the vessel to the tumor is still $d=0$ mm. Then vessel flow ranges in $y/L=$ [0.4,0.45] for $d=5$ mm, in $y/L=$ [0.35,0.45] for $d=10$ mm, and $y/L=$ [0.25,0.45] for $d=20$ mm. As illustrated in Figure \ref{FigEffectWidthCEMCT} (c), there shows a width of 5 mm from the upper blood vessel downwards. Obviously, the thinner the blood vessel, the larger the velocity gradient near the upper vessel wall (adjacent to the tumor). In Figure \ref{FigEffectWidthCEMCT} (a), almost all the contour lines on the MNP concentrations are overlapped except for a little part much near the vessel. This is consistent with the analysis in Section \ref{SecEffectVessel}. However, the isotherms obtain a large divergence among the three cases not only significantly performing near the blood vessel but also the upstreaming region  of the tumor (see Figure \ref{FigEffectWidthCEMCT} (b)). As the width of the vessel extends, the area of the same isotherm expands, and the maximum temperature elevates as Table \ref{TableEffectWidth}. This change surely results in the difference to tumor treatment efficacy. Regarding the Figure \ref{FigEffectWidthCEMCT} (d), a considerable benefit is added when the width of the blood vessel converts from 5 mm to 20 mm, which has been quantified in Table \ref{TableEffectWidth}.

\begin{table}
    \centering
    \caption{Effect of blood vessel width on maximum temperature and $R_{CEM43}$ in tumor at treatment time $t=60$ min.}
    \begin{tabular}{cccc}
    \toprule
         $w$, mm & 5 & 10 & 20  \\ \midrule
         $T_{max}$, $^\circ C$ & 43.93 & 44.19 & 44.39   \\
         $R_{CEM43}$ in tumor, \% & 57.32 & 67.27 & 74.76   \\ \bottomrule
    \end{tabular}
    \label{TableEffectWidth}
\end{table}

\subsection{Effect of vessel direction}
\begin{figure}
    \centering
    \includegraphics[width=\textwidth]{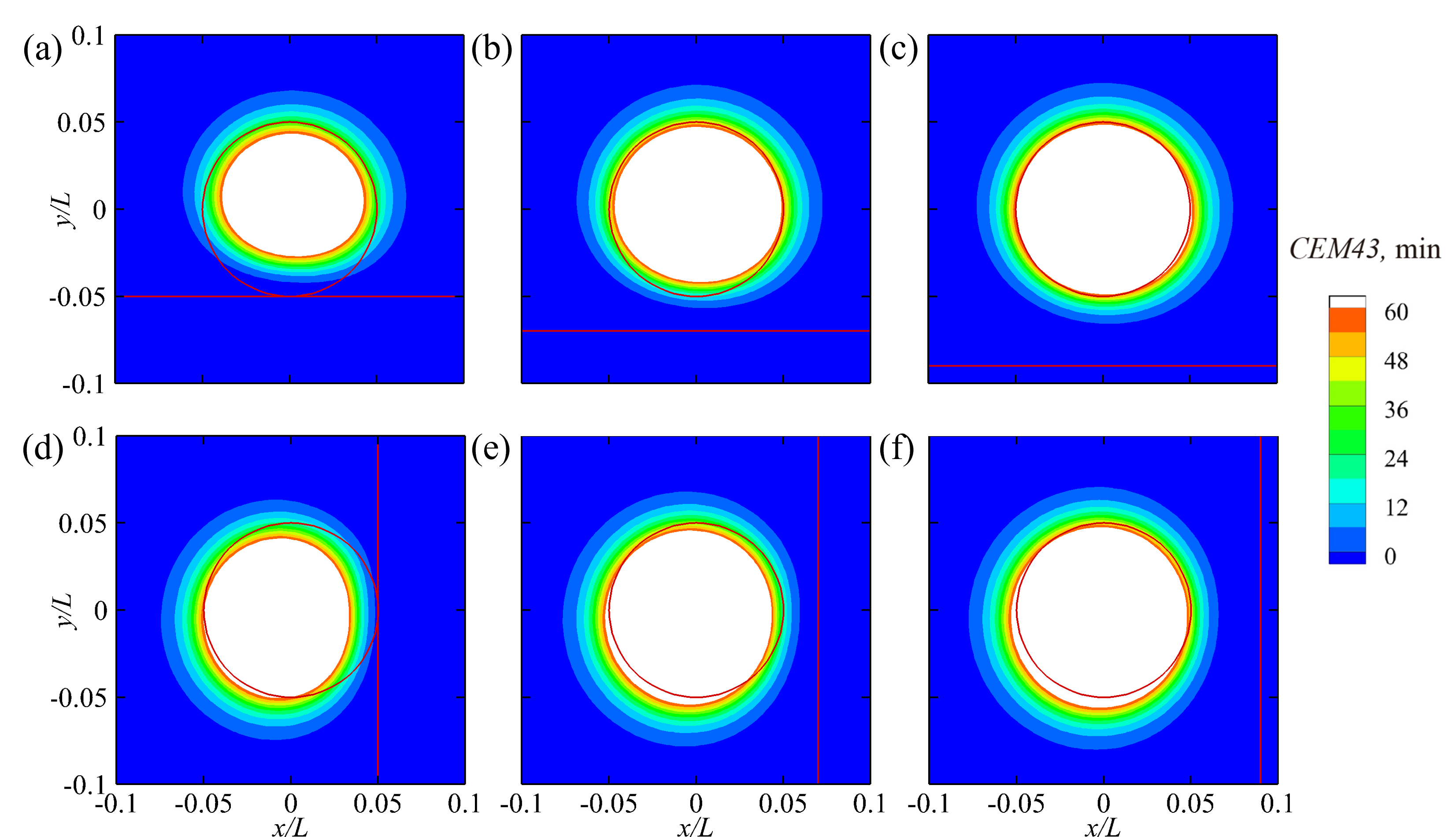}
    \caption{Effect of vessel direction of blood vessel on the CEM43 distribution at treatment time $t=60$ min. Horizontal blood vessel with distance (a) $d=0$ mm, (b) $d=2$ mm, and (c) $d=4$ mm. Downwards vertical blood vessel with distance (a) $d=0$ mm, (b) $d=2$ mm, and (c) $d=4$ mm. The red circle means the tumor boundary, while the red horizontal/vertical line denotes the blood vessel boundary towards the tumor side.}
    \label{FigEffectDirection}
\end{figure}
Figure \ref{FigEffectDirection} displays the effect of blood vessel direction on treatment efficacy. As discussed in Section \ref{EffectGravity}, the collaboration of downward gravity and the proper cooling effect of the blood vessel from the lower part makes the horizontal vessel with $d=4$ mm the best performance among the horizontal cases. However, the downward vertical vessel cools down the temperature from the right part, which cannot cancel the side effects of gravity, causing an easy-ablation region to the southwest of the tumor and a hard-ablation region in the opposite tumor domain. $R_{CEM43}$ at $t=60$ min treatment time can be found in the Table \ref{TableEffectDirection}, a similar consequence can be drawn with which.

\begin{table}
    \centering
    \caption{Effect of blood vessel direction on $R_{CEM43}$, \%}    
    \begin{tabular}{cccc}
    \toprule
         $d$, mm & Position & Horizontal & Vertical \\ \midrule
         \multirow{2}{*}{4} & Tumor & 97.77 & 96.18 \\
          & Healthy tissue & 0.87 & 8.75    \\
         \multirow{2}{*}{2} & Tumor & 85.52 & 90.37 \\
          & Healthy tissue & 0.00 & 5.49    \\
         \multirow{2}{*}{0} & Tumor & 57.32 & 75.42 \\
          & Healthy tissue & 0.00 & 1.51    \\ \bottomrule
    \end{tabular}
    \label{TableEffectDirection}
\end{table}

\subsection{Effect of moving vessel boundary}
\begin{figure}
    \centering
    \includegraphics[width=\textwidth]{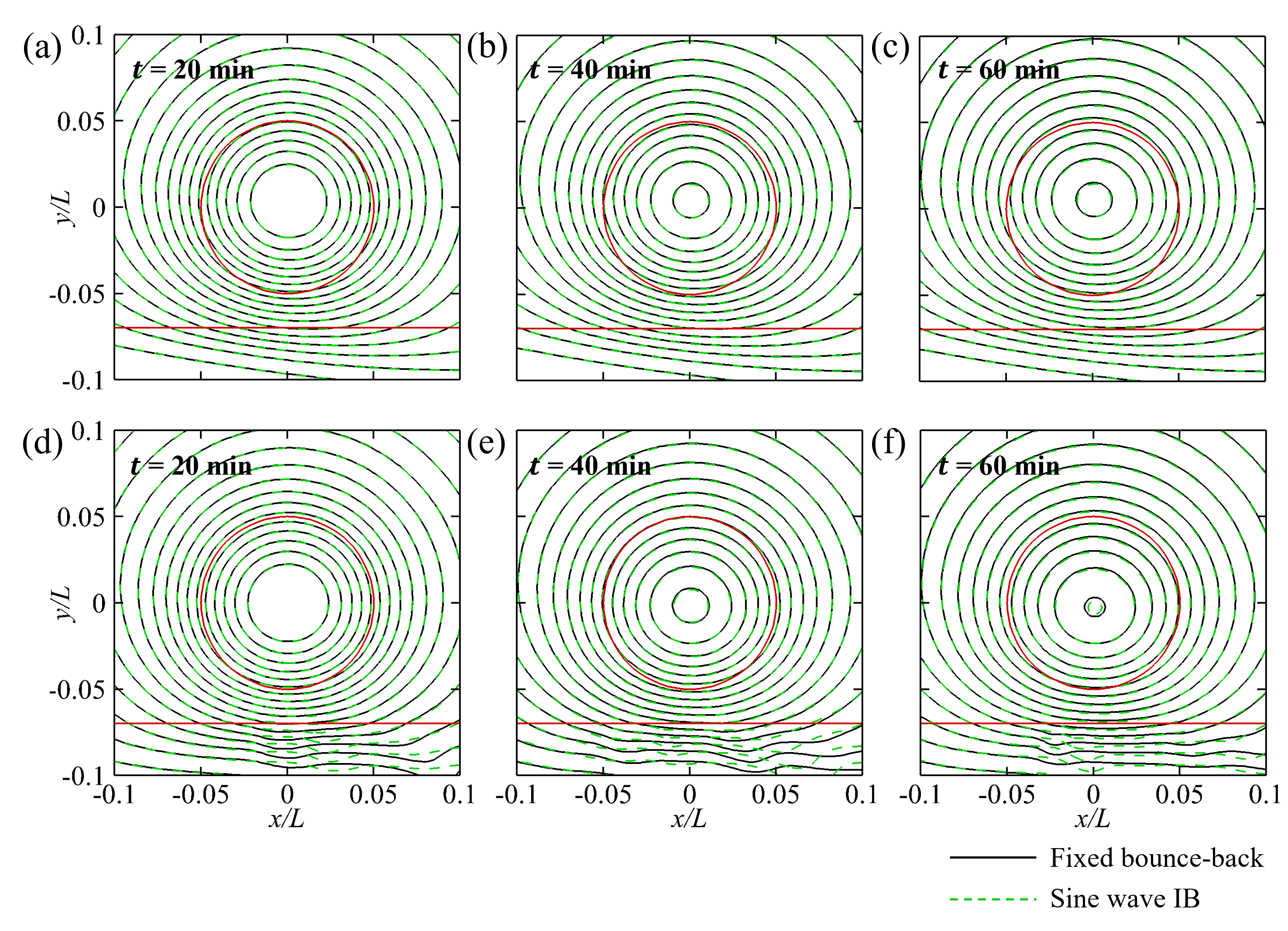}
    \caption{Effect of moving vessel on the isotherms. (a) $\sim$ (c) denote the isotherms with no gravity effect. (d) $\sim$ denote the isotherms with the gravity effect.}
    \label{FigEffectMovingVessel}
\end{figure}

In this part, the moving boundary of blood vessel walls is involved to explore the influence of which. The width of blood vessel is $w=10$ mm, and the variation of vessel width is $\Delta w=0.04w$. The results of traditional bounce-back, fixed IB, sine wave IB (wavelength $\lambda=L$, frequency $f=1$ Hz, and phase difference $\Delta \phi=\pi /4$ ), and pulsation wave IB (referred to \cite{cinthio2006longitudinal}) are listed in Table \ref{TableEffectMovingVessel} for comparison. Results indicate that, for the simulated case, there exists only a little variation on $R_{CEM43}$ in tumor and healthy tissue for four different vessel boundaries, even the moving IB boundaries, which means moving boundary does not bring a significant influence on magnetic hyperthermia treatment. In addition, we compare the isotherms of fixed IB and sine wave IB as Figure \ref{FigEffectMovingVessel}. Figures confirm there is almost no impact on the treatment efficacy.

\begin{table}
    \centering
    \caption{Effect of moving blood vessel wall on $R_{CEM43}$ in tumor and healthy tissue at treatment time $t=60$ min.}
    \begin{tabular}{ccccccc}
    \toprule
         $d$, mm & Condition & Position & Fixed BB & Fixed IB & Sine IB & Pulsation IB \\   \midrule
         \multirow{4}{*}{2} & \multirow{2}{*}{No gravity} & Tumor & 88.78 & 89.50 & 89.58 & 89.82   \\
         & & Healthy tissue & 1.83 & 2.23 & 1.91 & 1.59    \\
         & \multirow{2}{*}{Gravity} & Tumor & 93.48 & 93.72 & 93.72 & 92.84 \\
         & & Healthy tissue & 0.32 & 0.24 & 0.16 & 0.08   \\
         \multirow{2}{*}{0} & \multirow{2}{*}{No gravity} & Tumor & 67.28 & 67.13 & 68.81 & -   \\
         & & Healthy tissue & 0.00 & 0.00 & 0.00 & -  \\
         \bottomrule
    \end{tabular}
    \label{TableEffectMovingVessel}
\end{table}

\subsection{Effect of blood rheology}
Blood is a shear-rate-dependent non-Newtonian fluid. In this part, Carreua-Yasuda Model is applied to study the influence of blood rheology (see \ref{BloodRheology} for details). Values of the parameter in Carreua-Yasuda Model are $\eta_0=0.056$ Pa$\cdot$s, $\eta_\infty=0.00345$ Pa$\cdot$s, $\lambda=3.313$ s, $n=0.22$, $a=1.25$ \cite{cho1991effects}. The comparison with Newtonian on $R_{CEM43}$ is listed in the table \ref{TableEffectRheology}. Since blood is regarded as a shear-thinning viscous flow, the velocity gradient near the vessel wall behaves a little larger than that of non-Newtonian. Not surprisingly, the degree of ablation in tumor and healthy tissue, therefore, is a bit more moderate. This is consistent with the effect of the vessel width.
\begin{table}
    \centering
    \caption{Effect of blood rheology on $R_{CEM43}$ (Unit: \%) at treatment time $t=60$ min}
    \begin{tabular}{cccc}
    \toprule
         $d$, mm & Position & Newtonian & Carreua-Yasuda Model \\ \midrule
         \multirow{2}{*}{2} & Tumor & 89.50 & 89.10 \\
          & Healthy tissue & 2.23 & 1.83    \\
         \multirow{2}{*}{0} & Tumor & 69.13 & 67.06 \\
          & Healthy tissue & 0.00 & 0.00    \\ \bottomrule
    \end{tabular}
    \label{TableEffectRheology}
\end{table}

\section{Conclusion}
This study developed a holistic simulation framework for magnetic hyperthermia cancer treatment and investigated the situations when the tumor is near a blood vessel. Simulation involves the blood and interstitial tissue flow, MNP distribution, temperature profile, and the behaviors of nanofluids. The treatment efficacy is evaluated with the thermal dose CEM43 model. We investigate the influence of the nearby blood vessel, the distance from the vessel to the tumor, the width of the vessel, the direction of the vessel, the moving vessel wall, and the blood rheology. 

According to the results, the nearby vessel has a significant cooling effect on the treatment. A vessel next to the tumor can reduce the ablation areas to less than half. Gravity creates a downward tendency of temperature profile and a little reduction in the maximum temperature, but these changes seem much more limited compared with the influence of the distance. Although narrow, the downward influence combines the upward cooling effect from the vessel, resulting in the best performance in treatment at distance $d=4$ mm. A smaller distance from the tumor to the vessel cools down the temperature a lot, therefore considerably prolonging the treatment procedure. However, the effect on temperature almost vanishes as distance expends to 8 mm. Width the same mean velocity, divergence widths of the blood vessel cause the different velocity gradient near the vessel wall. The thinner the blood vessel, the lower the temperature profile, and the slower the tumor ablation pace. Not fortunate as the horizontal vessel, the downward vertical vessel cools the temperature from the right part, which cannot reduce the gravity effect on treatment, so the performance is inferior to the horizontal blood vessel. Blood rheology leads to a little larger gradient near the vessel wall, then resulting in a bit more moderate ablation process. Moving vessel boundaries exert the bare impact on the treatment efficacy. 

\section*{Acknowledgement}
This study was financially supported by the Research Grants Council of Hong Kong under General Research Fund (Project No. 15214418). 

\appendix
\section{Lattice Boltzmann method}
The LBM is a relatively new CFD method for fluid flow and heat/mass transfer simulations. Unlike traditional CFD methods, which solve the conservation equations of macroscopic properties numerically, LBM models the fluid particles by distribution functions through consecutive streaming and collision processes over a number of square lattices \cite{wang2022simulation,ren19,wang2016control}. Zhang \cite{zhang2008lattice} was probably the first to apply the LBM to solving PBHTE, successfully demonstrating the capability of LBM in simulating bioheat problems. This mesoscopic approach was then widely applied for bioheat studies \cite{golneshan2011diffusion,das2013estimation,das2013numerical}. In the present study, a D2Q9 (i.e., two-dimensional nine-discrete-velocity) MRT (i.e., multiple-relaxation-time, a collision model that is used to improve the numerical stability \cite{lallemand2000theory}) and a D2Q5 (i.e., two-dimensional five-discrete-velocity) MRT LBM scheme are namely applied to obtain the vector field (flow field)  and scalar field (temperature field and concentration field) by solving Equation \ref{EqnGoverning} \cite{liu2014multiple,liu2018multiple}.

The discrete D2Q9 MRT-LBM equation for velocity field is written as
\begin{equation}
    \mathbf{f}(x_k+\mathbf{e}\delta_t,t_n+\delta_t)-\mathbf{f}(x_k,t_n)=-\mathbf{M}^{-1} \mathbf{\Lambda}[\mathbf{m}-\mathbf{m}^{(eq)}]|_{(x_k,t_n)}+\mathbf{M}^{-1}\delta_t(1-\frac{\mathbf{\Lambda}}{2})\mathbf{S}
    \label{EqnD2Q9}
\end{equation}
where $\mathbf{f}(x_k,t_n)$ is nine-dimensional distribution function vectors at time ${t}_n$ and node $x_k$ for fluid field. $\mathbf{m}$ and $\mathbf{m}^{(eq)}$ are moment and the corresponding equilibrium moment vector for flow field. $\mathbf{e}$ describes unit velocities along 9 discrete directions
\begin{equation}
    e_i=
    \begin{cases}
            (0,0) & {i=0} \\
            (\cos{[(i-1)\pi/2]},\sin{[(i-1)\pi/2]})c  & {i=1\sim 4} \\
            (\cos{[(2i-9)\pi/4]},\sin{[(2i-9)\pi/4]})\sqrt{2}c  & {i=5\sim 8}
    \end{cases}
\end{equation}
where $c=\delta_x/\delta_t$ is the lattice speed, which is 1 since $\delta_x=\delta_t$ in the MRT model. 
$\mathbf{M}$ is a $9\times5$ orthogonal transformation matrix
\begin{equation}
\mathbf{M}=
    \begin{pmatrix} 
    1&1&1&1&1&1&1&1&1 \\ 
    -4&-1&-1&-1&-1&2&2&2&2 \\
    4&-2&-2&-2&-2&1&1&1&1 \\
    0&1&0&-1&0&1&-1&-1&1 \\
    0&-2&0&2&0&1&-1&-1&1 \\
    0&0&1&0&-1&1&1&-1&-1 \\
    0&0&-2&0&2&1&1&-1&-1 \\
    0&1&-1&1&-1&0&0&0&0 \\
    0&0&0&0&0&1&-1&1&-1
    \end{pmatrix}
\end{equation}
$\mathbf{\Lambda}$ is the nine-dimensional diagonal relaxation matrix
\begin{equation}
    \mathbf{\Lambda}=\text{diag}(1,1.1,1.1,1,1.2,1,1.2,1/\tau_\upsilon,1/\tau_\upsilon)
\end{equation}
and $\tau_\upsilon$ can be recovered to viscosity of nanofluid in Chapman–Enskog analysis on Equation \ref{EqnGoverning} as
\begin{equation}
    \upsilon_{nf}=\sigma c_{s}^2(\tau_\upsilon-0.5)\delta_t
\end{equation}
$\mathbf{S}$ is the external force vector in the moment space, which is linked to the body force $F$ in governing equation \ref{EqnGoverning}(b).

The discrete D2Q5 MRT-LBM equation for temperature and concentration fields are written as
\begin{subequations}
    \begin{equation}
        \mathbf{g}(x_k+\mathbf{e}\delta_t,t_n+\delta_t)-\mathbf{g}(x_k,t_n)=-\mathbf{N}^{-1} \mathbf{\Theta}[\mathbf{n_g}-\mathbf{n_g}^{(eq)}]|_{(x_k,t_n)}+\mathbf{N}^{-1}\delta_t\mathbf{\Psi}
    \end{equation}
    \begin{equation}
    \mathbf{h}(x_k+\mathbf{e}\delta_t,t_n+\delta_t)-\mathbf{h}(x_k,t_n)=-\mathbf{N}^{-1} \mathbf{\Upsilon}[\mathbf{n_h}-\mathbf{n_h}^{(eq)}]|_{(x_k,t_n)}
\end{equation}
\end{subequations}
where $\mathbf{g}(x_k,t_n)$ and $\mathbf{h}(x_k,t_n)$ are five-dimensional distribution function vectors at time ${t}_n$ and node $x_k$ for temperature and concentration respectively. $\mathbf{n}$ and $\mathbf{n}^{(eq)}$ are moment and the corresponding equilibrium moment vector, respectively, where subscribe "g" represents temperature field and "h" denotes concentration field. $\mathbf{e}$ describes unit velocities along 5 discrete directions
\begin{equation}
    e_i=
    \begin{cases}
            (0,0) & {i=0} \\
            (\cos{[(i-1)\pi/2]},\sin{[(i-1)\pi/2]})c  & {i=1\sim 4} \\
    \end{cases}
\end{equation}
$\mathbf{N}$ is a $5\times5$ orthogonal transformation matrix
\begin{equation}
\mathbf{N}=
    \begin{pmatrix} 
    1&1&1&1&1 \\ 
    0&1&0&-1&0 \\
    0&0&1&0&-1 \\
    -4&1&1&1&1 \\
    0&1&-1&1&-1
    \end{pmatrix}
\end{equation}
$\mathbf{\Theta}$ and $\mathbf{\Upsilon}$ are the diagonal relaxation matrix
\begin{subequations}
    \begin{equation}
        \mathbf{\Theta}=\text{diag}(1,1/\tau_T,1/\tau_T,1.5,1.5)
    \end{equation}
    \begin{equation}
        \mathbf{\Upsilon}=\text{diag}(1,1/\tau_C,1/\tau_C,1.5,1.5)
    \end{equation}
\end{subequations}
where $\tau_T$ can be linked to effective thermal diffusivity (in temperature field) or effective concentration diffusivity (in concentration field) in Chapman–Enskog analysis on Equation \ref{EqnGoverning} as
\begin{equation}
    \alpha_e=\sigma c_{sT}^2(\tau_T-0.5)\delta_t,\;D_e=\phi c_{sT}^2(\tau_C-0.5)\delta_t
\end{equation}

$\mathbf{\Psi}$ is a heat source vector, which can is connected with the heat source $Q$ in governing equation \ref{EqnGoverning}(c). More details about the D2Q5 MRT LBM can be found in \cite{liu2014multiple,liu2018multiple}.

For the boundary conditions at four sides of healthy tissue block in this study, stationary wall is applied for fluid field, constant values are used on thermal and solutal fields, as depicted in Figure \ref{FigSchematicVessel}. Here, halfway bounce-back is adopted for stationary wall, while anti-bounce-back scheme is adopted for constant temperature and concentration boundary. In addition, at the interface of healthy tissue and tumor, it is deemed the same velocity, same temperature and same concentration.

\begin{table*}[]
    \centering
    \caption{Sensitivity study on grid number}
    \begin{tabular}{ccccccc}
    \toprule
    \multirow{2}{*}{$NX\times NY$} & \multicolumn{2}{c}{$t=60$, min} & \multicolumn{2}{c}{$t=80$, min} \\
     & $R_{CEM43}$ in tumor, \% & Error, \% & $R_{CEM43}$ in tumor, \% & Error, \%   \\ \midrule
    $200\times 200$ & 68.67 & 1.39 & 75.63 & 2.22   \\
    $400\times 400$ & 67.28 & 0.83 & 73.41 & 0.75   \\
    $600\times 600$ & 66.45 & - & 72.66 & -   \\
    \bottomrule   
    \end{tabular}
    \label{TableMeshSensi}
\end{table*}

A sensitivity study of  grid number is conducted. The baseline case with gravity effect is considered as an example, where three sets of grid size are compared. $R_{CEM43}$ in tumor regions are recorded, and they are listed as Table \ref{TableMeshSensi}. Results suggest $400\times400$ grid number is suitable for the study.

\begin{figure*}[htpb]
    \centering
    \includegraphics[width=\textwidth]{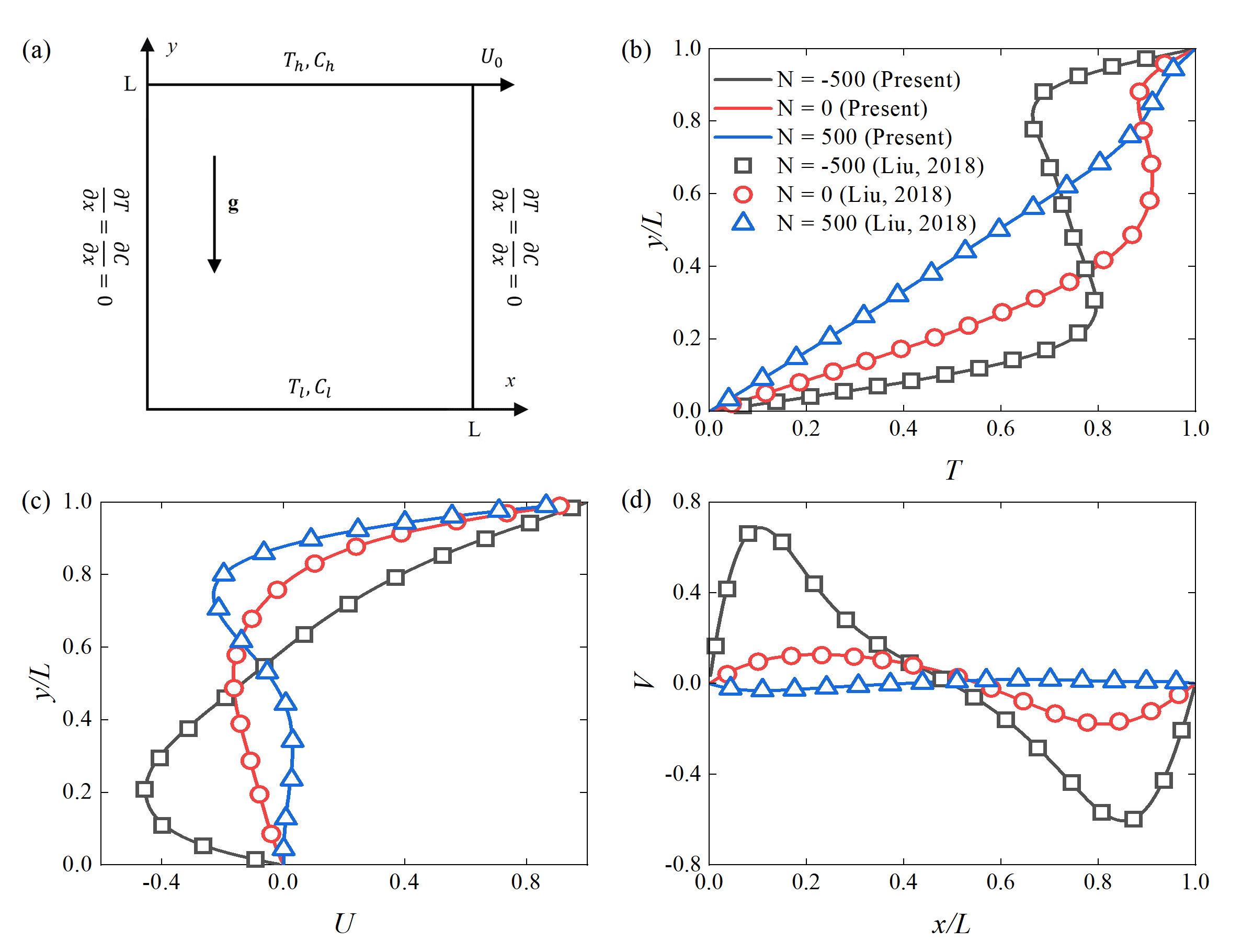}
    \caption{Validation of LBM framework on double-diffusive convection problem in porous media. (a) vertical temperature profile at x/L=0.5; (b) vertical velocity profile at x/L=0.5; (c) horizontal velocity profile at x/L=0.5.  
    (b) vertical velocity on $y=0.5$, (c) temperature on $y=0.5$}
    \label{FigDoubleDiff}
\end{figure*}

Since the magnetic hyperthermia problem governed by Equation \ref{EqnGoverning} actually can be viewed as a double-diffusive convection problem in porous media. Therefore, a typical porous double-diffusive convection validation is conducted to justify this framework. Figure \ref{FigDoubleDiff} show the comparison on velocity and temperature with Reference \cite{liu2018multiple} at various buoyancy ratio $N$. It is indicated that the present results match well with the reference.

\section{Implicit immersed boundary}
\label{AppendixImmersedBoundary}
Induced by the blood pulsation and deformable tissue domain, the vessel wall is regarded as a moving boundary. IBM is a mixed Eulerian–Lagrangian method by adding a force density term to satisfy the no-slip boundary condition. Generally, IBM can easily handle situations where the interface grids do not match in the Eulerian and Lagrangian configurations with desired numerical accuracy. In this research, a direct-forcing IBM scheme with the diffuse interface treatment is applied to simulate the movement of the blood vessel wall. 

Based on the Lattice Boltzmann Equation \ref{EqnD2Q9}, the boundary force in two-dimension at point $x_b$ is given by \cite{kang2010immersed}
\begin{equation}
    \mathbf{f}_b(s,t)=2\rho\frac{\mathbf{U}^d(s,t)-\mathbf{u}^{noF}(s,t)}{\delta t}
    \label{MethoEqnFb}
\end{equation}
where $\mathbf{U}^d(s,t)$ and $\mathbf{u}^{noF}(s,t)$ are namely expected and unforced velocity on the Lagrangian coordinate, and unforced velocity is interpolated from the Eulerian configuration as
\begin{equation}
    \mathbf{u}^{noF}(s,t)=\sum_k \mathbf{u}^{noF}(x_k,t)D(x_k-x_b(s,t))h^3
    \label{MethoEqnUnoFs}
\end{equation}
and the unforced velocity $\mathbf{u}^{noF}(s,t)$ is collected from the particle distribution function in LB Equation without forcing term as
\begin{equation}
    \mathbf{u}^{noF}(x_k,t)=\frac{1}{\rho}\sum_ie_if_i(x_k,t)
    \label{MethoEqnUnoFx}
\end{equation}

Then the boundary force on the Eulerian configuration can interpolate from the Lagrangian coordinate as
\begin{equation}
    \mathbf{F}_b(x_k,t)=\sum_b\mathbf{f}_b(s,t)D(x_k-x_b)\delta s_b
    \label{MethoEqnFx}
\end{equation}

Finally, velocity on Eulerian configuration updates by
\begin{equation}
    \mathbf{u}(x_k,t)=\mathbf{u}^{noF}(x_k,t)+\frac{\delta t}{2\rho}\mathbf{F}_b(x_k,t)
    \label{MethoEqnUx}
\end{equation}

To better satisfy the no-slip boundary condition, a new velocity is obtained by interpolating from $\mathbf{u}(x_k,t)$ with Equation \ref{MethoEqnUnoFs} to substitute $\mathbf{u}^{noF}(s,t)$, and then iterating the Equation \ref{MethoEqnFb}, \ref{MethoEqnFx}, \ref{MethoEqnUx}, \ref{MethoEqnUnoFs} until the difference between the updated velocity and the desired velocity is small enough. 

In this process, $x_k$ is the independent Cartesian coordinates, while $s$ is the independent curvilinear coordinate. $h=\delta x_k$ denotes the meshing space and $\delta s_b$ is the segment length of boundary. 

Delta function applied on interpolation is given by
\begin{equation}
    D(x_k-x_b)=\frac{1}{h^2}d(\frac{x_k-x_b}{h})d(\frac{y_k-y_b}{h})
\end{equation}
where $d(r)$ is a weighting function defined as
\begin{equation}
    d(r)=
    \begin{cases}
            \frac{1}{8}(3-2|r|+\sqrt{1+4|r|-4r^2}) & |r|\leq1 \\
            \frac{1}{8}(5-2|r|+\sqrt{-7+12|r|-4r^2}) & 1\leq|r|\leq2 \\
            0 & |r|\ge2
    \end{cases}
    \label{MethoEqnDr}
\end{equation}

The cases of flow past the cylinder are applied on the validation of immersed boundary. The uniform flow past a fixed circular cylinder is studied. Flow is coming from the left boundary with the velocity $U_\infty$, and the homogeneous Neumann boundary is assigned at the right boundary. We compare two steady flow conditions ($Re = 20$ and $40$, with $40D\times 40D$ computational domain) and two vortex shedding flow conditions ($Re = 100$ and $150$, with $60D\times 40D$ computational domain). The results are listed in Table \ref{TabValidKang2010}, and obviously match well with the reference.

\begin{table}
    \centering
    \caption{Validation of immersed boundary method on flow past cylinder case in \cite{kang2010immersed}}
    \setlength{\tabcolsep}{0.3cm}{
    \begin{tabular}{c|cccccc|cccccc}
        \toprule
        \multicolumn{13}{c}{Steady flow}  \\    \bottomrule
        Reynolds number & \multicolumn{6}{c|}{20} & \multicolumn{6}{c}{40}  \\   \hline
        Parameters & \multicolumn{3}{c}{$C_D$} & \multicolumn{3}{c|}{$L_w$} & \multicolumn{3}{c}{$C_D$} & \multicolumn{3}{c}{$L_w$} \\
        Kang, 2010 & \multicolumn{3}{c}{2.119} & \multicolumn{3}{c|}{1.033} & \multicolumn{3}{c}{1.584} & \multicolumn{3}{c}{2.486} \\
        Present &  \multicolumn{3}{c}{2.137} & \multicolumn{3}{c|}{1.036} & \multicolumn{3}{c}{1.597} & \multicolumn{3}{c}{2.508} \\ \toprule
        \multicolumn{13}{c}{Unsteady flow}  \\  \bottomrule
        Reynolds number & \multicolumn{6}{c|}{100} & \multicolumn{6}{c}{150}  \\ \hline
        Parameters & \multicolumn{2}{c}{Avg. $C_D$} & \multicolumn{2}{c}{$C_L$} & \multicolumn{2}{c|}{$St$} & \multicolumn{2}{c}{Avg. $C_D$} & \multicolumn{2}{c}{$C_L$} & \multicolumn{2}{c}{$St$} \\
        Kang, 2010 & \multicolumn{2}{c}{1.368} & \multicolumn{2}{c}{$\pm0.346$} & \multicolumn{2}{c|}{0.162} & \multicolumn{2}{c}{1.351} & \multicolumn{2}{c}{$\pm0.542$} & \multicolumn{2}{c}{0.181} \\
        Present & \multicolumn{2}{c}{1.381} & \multicolumn{2}{c}{$\pm0.355$} & \multicolumn{2}{c|}{0.164} & \multicolumn{2}{c}{1.367} & \multicolumn{2}{c}{$\pm0.543$} & \multicolumn{2}{c}{0.181} \\
        \toprule
    \end{tabular}    } 
    \label{TabValidKang2010}
\end{table}

\section{Blood rheology}
\label{BloodRheology}
Blood is a complex mixture of erythrocytes, leukocytes, thrombocytes, iron oxides, and other components suspended in plasma \cite{tzirtzilakis2005mathematical,wang2011lattice}, which makes the shear rate of blood fluid is not linearly related to the viscosity. Blood rheology can be regarded as a shear-rate-dependent non-Newtonian fluid \cite{cho1991effects}.  Already some experiments and numerical studies reveal the rheological behavior of blood, and several shear-thinning models are accepted in the related works, such as Power-Law Model \cite{loenko2019natural}, Casson Model \cite{walawender1975approximate}, and Carreua-Yasuda Model \cite{kefayati2019three, kefayati2014fdlbm, reza18-1, reza18-2, kim2016spectral}. Since Carreua-Yasuda can describe a wild scope of blood flow and has been commonly applied for portraying the shear-shinning behavior of blood, which is selected in this research. 

Generally, the Carreua-Yasuda Model is defined as
\begin{equation}
    \eta=\eta_\infty+(\eta_0-\eta_\infty)[1+(\lambda \dot{\gamma})^a]^{\frac{n-1}{a}}
    \label{EqnCYModel}
\end{equation}
where $\eta_0$ and $\eta_\infty$ are the viscosity at the zero shear rate and infinite shear rate, respectively. $n$ denotes the power-law exponent about the flow behavior. $a$ is the parameter controlling the width of the transition region. $\lambda$ is a time constant, whose inverse $\frac{1}{\lambda}$ denotes the position where the viscosity start to drop. 

Shear rate $\dot\gamma$ is a relevant parameter to the second invariant of the rate-of-strain tensor and is defined as 

\begin{equation}
        \dot\gamma=\sqrt{\frac{1}{2}\Pi}=\sqrt{\frac{1}{2}[\sum_i\sum_j\dot\gamma_{ij}\dot\gamma_{ji}]}
        \label{EqnGamma}
\end{equation}
and $\dot\gamma_{ij}$ is defined as

\begin{equation}
    \dot\gamma_{ij}=u_{i,j}+u_{j,i}
    \label{EqnGammaij}
\end{equation}

\begin{figure}
    \centering
    \includegraphics[width=\textwidth]{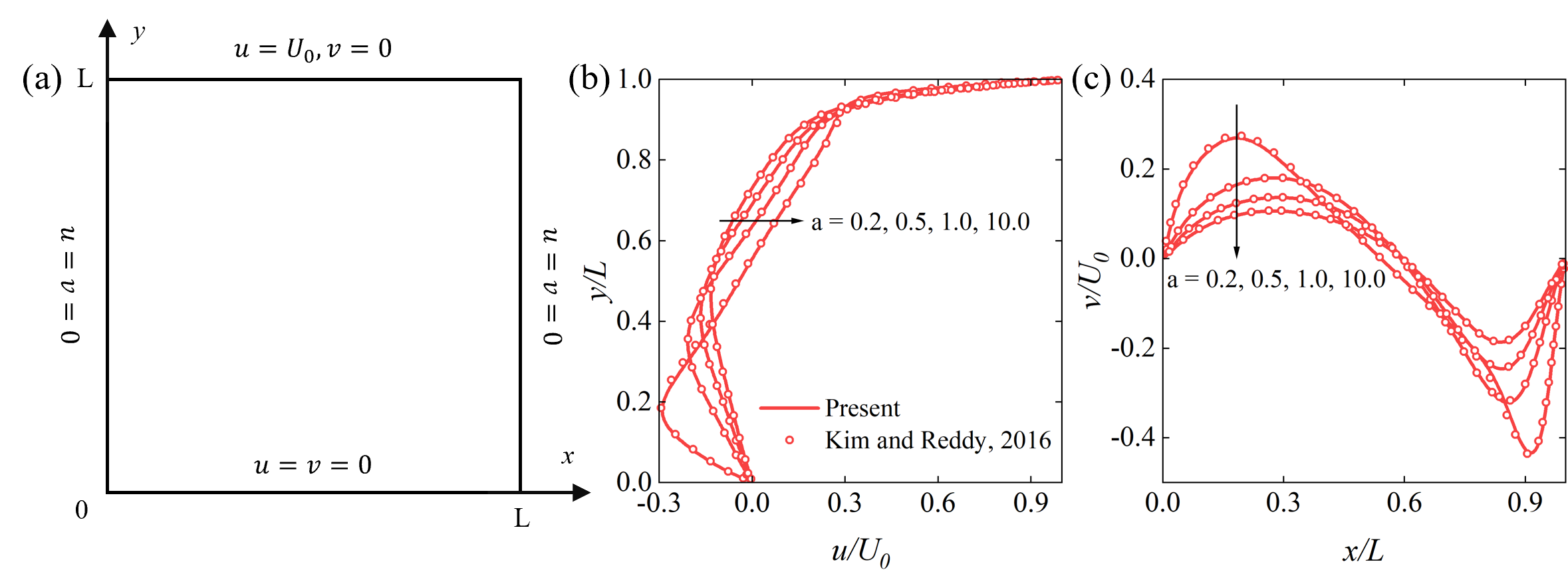}
    \caption{Validation of Carreua-Yasuda non-Newtonian Model on a lid-driven cavity flow field. (a) Schematic in \cite{kim2016spectral}; (b) Horizontal velocity at $x/L=0.5$, and (c) Vertical velocity at $y/L=0.5$ with various values of $a$.}
    \label{FigValidnonNewtonian}
\end{figure}

To validate the blood rheology with Carreua-Yasuda Model, a lid-driven cavity flow field is adopted for comparison. As Figure \ref{FigValidnonNewtonian} (a) illustrates, the upper wall moves with a constant velocity $U_0$, while the other three walls are fixed. Figure \ref{FigValidnonNewtonian} (b) and (c) present the horizontal velocity at $x/L=0.5$ and vertical velocity at $y/L=0.5$, respectively, where the parameter $a$ (see Equation \ref{EqnCYModel} varies from 0.2 to 10.0. According to the results, the velocity profiles match well with the reference.

 \bibliographystyle{elsarticle-num} 
 \bibliography{ref}





\end{document}